\documentclass[preprint,aps,amsmath]{revtex4}
\usepackage{graphicx}
\usepackage{amssymb}
\usepackage{bm}

\begin{document}

%\linenumbers

\title{Gyrofluid vortex interaction}

\author{Alexander Kendl}
\affiliation{Institut f\"ur Ionenphysik und Angewandte Physik, Universit\"at
  Innsbruck, 6020 Innsbruck, Austria \vspace{2cm}
} 

\begin{abstract} \vspace{1cm}
Low-frequency turbulence in magnetised plasmas is intrisically influenced by
gyroscale effects across ion Larmor orbits.   
Here we show that fundamental vortex interactions like merging and
co-advection in gyrofluid plasmas are essentially modified under the influence
of gyroinduced vortex spiraling. 
For identical initial vorticity, the fate of co-rotating eddies is decided
between accelerated merging or explosion by the asymmetry of initial density
distributions. 
Structures in warm gyrofluid turbulence are characterised by gyrospinning
enhanced filamentation into thin vorticity sheets.   
\end{abstract} 

\maketitle

\section{Introduction}

Vortices can be regarded as the basic constituents of turbulence. 
Vortex motion and interactions govern nonlinear structure formation, flow
and convective transport properties in a variety of fluids.
A particular case of interest are quasi-two-dimensional fluids, which are
characterised by the possibility for formation of coherent structures and
large scale (zonal) flows, a dual cascade, and the ideal conservation of
enstrophy in addition to energy \cite{kraichnan80,boffetta12}.  
Merging and filamentation of vortices are fundamental processes that underly
these properties. 
Examples for quasi-2D fluids encompass stratified and rotating fluids,
atmospheric and oceanic flows, or the cross-field dynamics of magnetised
plasmas \cite{horton94}. 

In magnetised space, laboratory or fusion plasmas, the fluidlike convection
perpendicular to a magnetic field ${\bf B}$ is governed by drifts, which
describe the mean motion on top of fast charged particle gyration. 
In particular, the magnetic confinement of fusion plasmas is crucially
determined through turbulent transport generated by drift-type
instabilities, and their suppression by (turbulence driven) zonal or equilibrium
flows \cite{horton99,scott10cpp}.  
Turbulent convection in magnetised plasmas is dominated by the
``E-cross-B'' drift velocity 
${\bf v}_E = {\bf E} \times {\bf B} / B^2$ 
%${\bf v}_E = (1/B^2) \; {\bf E} \times {\bf B}$ 
in presence of a fluctuating electric field ${\bf E} = - {\bf \nabla} \phi$. 
A localised electric potential $\phi({\bf x}, t)$ leads to vortical
E$\times$B drift motion of the plasma around the potential perturbation. 
The vorticity ${\bf \Omega} = {\bf \nabla} \times {\bf v}$ of E$\times$B flows
can be expressed as ${\bf \Omega} = ({\bf B}/B^2) \nabla_{\perp}^2 \phi$ for
constant ${\bf B}$. 
The electric potential $\phi$ thus has the role 
of a stream function for E$\times$B flows.

Various instabilities driven by pressure gradients in magnetised plasmas
typically result in drift vortex structures with sizes around the drift scale
$\rho_0  = \sqrt{m_i T_e} /(e B)$, where $m_i$ is the ion mass, $T_e$ the
electron temperature, and $e$ the elementary charge.
For warm ion plasmas with non-zero temperature ratio $\tau_i = T_i / T_e \sim
1$, the drift wave and vortex scales are in the order of $\rho_i =
\sqrt{\tau_i} \; \rho_0$, the ion finite Larmor radius (FLR).  
Over the fast gyration of ions with particle charge $q_i = Z e$ around their
gyrocenters (at cyclotron frequency $\omega_i = q_i B / m_i$) they effectively 
experience a ring-averaged electric potential $\phi_i$ rather than the
potential at the gyrocenter (for low-frequency perturbations with $\omega
\ll \omega_i$), and contribute to a quasi-neutral spatial distribution with
approximately equal electron and ion particle densities $N_e ({\bf x}) \approx
N_i ({\bf x})$ on average over the gyroorbit.  

The framework for a description of magnetised plasmas governed by drift motion
with FLR effects is efficiently given by gyrokinetic models evolving a 5D
distribution function in phase-space \cite{hahm88,brizard07,krommes12},  
or by gyrofluid models of appropriate respective fluid moments in 3D space
\cite{knorr88,dorland93,beer96,scott05}.  
In the following, the effects of finite Larmor orbits on fundamental vortex
interactions are analysed within an isothermal gyrofluid model \cite{scott03}.
It is found that spiraling of vorticity induced by FLR effects in the presence
of spatial asymmetry significantly alters the merger process and generates
fine structured vorticity sheets. Specific initial conditions result either
in strongly accelerated merging or in vortex explosion, with consequences for
inverse turbulent cascade properties and zonal flows.

\section{Gyrofluid model for basic vortex interactions}

Here we employ a basic local (``delta-f'') gyrofluid model, derived from an
energetically consistent gyrofluid electromagnetic model \cite{scott10}
in the limit of 2D isothermal dynamics in a straight and constant magnetic
field. Details on the normalisation of the model equations and their numerical
implementation can be found in Ref.~\cite{kendl15}. 
Gyrofluid models do not dynamically evolve the particle densities $N_s({\bf
  x}, t)$ of electrons and ions ($s\in e,i$), but rather their gyrocenter
densities $n_s({\bf x}, t)$. For small density fluctuation amplitudes both are
connected through polarisation \cite{pfirsch84,dorland93,scott03,brizard13} 
of gyroorbits by $\phi$:

\begin{equation}
N_s = \Gamma_{1s} n_s + {\mu_s \over \tau_s} (\Gamma_{0s} -1) \phi.
\label{e:densities}
\end{equation}

The operator $\Gamma_{1s} = \Gamma_{0s}^{1/2}$ models gyroaveraging of the
gyrocenter density, and $\Gamma_{0s}$ reflects the gyroscreening
of the potential $\phi$ in the mass dependent ($\mu_s \equiv m_s/m_i$)
polarisation term. 
In wavenumber space $\Gamma_0(b_s) = I_0(b_s) \exp(-b_s)$ is expressed by the
modified Bessel function $I_0$ with $b_s = \rho_s^2 k_{\perp}^2 = \tau_s \mu_s
(\rho_0 k_{\perp})^2$. For electrons, $\mu_e \ll 1$, 
and the gyro radius $\rho_e \ll \rho_i$ is negligible, 
so that $b_e \approx 0$ and $\Gamma_{0e} = \Gamma_{1e} \equiv 1$ can be
assumed at ion gyro scales. The small electron mass also leads
to negligible polarisation, so that particle and gyrocenter densities 
$N_e \approx n_e$ coincide. The electron and ion particle densities are
connected by strict quasi-neutrality to $N_i \equiv N_e$ at scales much larger
than the Debye length.  

In the absence of driving, damping or parallel coupling the evolution equation
for the gyrocenter densities in a homogeneous magnetic field then reflects
incompressible mass conservation, and can (for normalised $B=1$) in 2D be
written as 

\begin{equation} 
\partial_t n_s + [ \phi_s , n_s] = 0
\label{e:continuity}
\end{equation}
where the advection term ${\bf v}_E   \cdot {\bf \nabla} n$
is expressed in Poisson bracket notation by $[\phi,n] = (\partial_x \phi)
(\partial_y n) - (\partial_y \phi)(\partial_x n)$,
describing advection of electron density along contours of the potential
$\phi$, and of ion gyrocenter density along the gyroaveraged potential
$\phi_i =  \Gamma_{1i} \phi$.   
For numerical stability a hyperviscosity term $- \nu_4 \nabla^4 n_s$ is
added to the right hand side of eq.~(\ref{e:continuity}).

The continuity eqs.~(\ref{e:continuity}) are closed by the quasi-neutrality
condition $N_i=N_e$, which by the gyrodensity relation in
eq.~(\ref{e:densities}) gives the polarisation equation: 

\begin{equation}
{1 \over \tau_i} (\Gamma_{0i} -1) \phi = n_e - \Gamma_{1i} n_i.
\label{e:polarisation}
\end{equation}
The gyro-operators can be written in the Pad\'e approximate forms
$\Gamma_0 = [1+b]^{-1}$ and $\Gamma_1 = [1+(1/2) b]^{-1}$ and using $b_i = -
\tau_i \mu_i (\rho_0 \nabla_{\perp})^2$ in real space \cite{dorland93}.

\section{Vortex merging for cold ions}

In the cold ion limit $\tau_i=0$ the polarisation equation reduces to
$\nabla_{\perp}^2 \phi = n_e - n_i$, which for normalised $B=1$ describes the
deviation between electron and ion gyrocenter densities in relation to a
scalar E$\times$B vorticity $\Omega = \nabla_{\perp}^2 \phi$.
Subtracting the ion from the electron gyrocenter density equation,
eq.~(\ref{e:continuity}) in this cold limit transforms into the classical 2D
Euler equation in vorticity representation:

\begin{equation}
d_t \Omega \equiv \partial_t \Omega + [\phi,  \Omega] = 0  \qquad
(\mbox{for}~\tau_i = 0). \label{e:vorticity}
\end{equation}

%%----------------------------------------
\begin{figure} % FIG 1
%%\vspace{0.5cm}
\includegraphics[width=12.0cm]{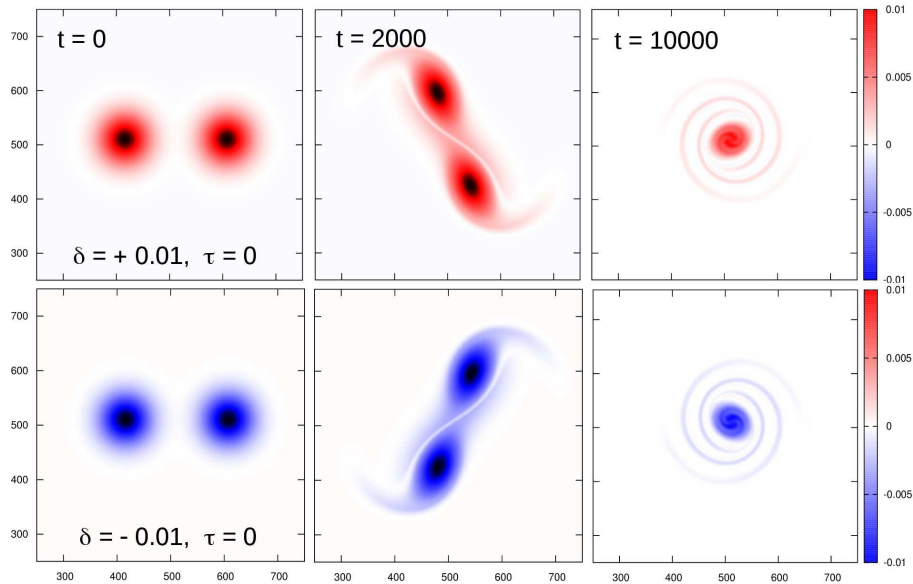}
\caption{\sl Evolution of the vorticity $\Omega({\bf x})$ field at several
  times during the co-rotating vortex pair interaction for $\tau_i=0$:
  anti-symmetric with the sign of the vorticity amplitude $\delta$ (top/red:
  positive, bottom/blue: 
  negative). }  
\label{f:merger0}
\end{figure}

Vortex interactions, in particular merging and co-advection, are in this limit
akin to classical fluids, and the gyrocenter densities are just passively
advected. 
For $\tau_i = 0$, a vortex that is defined by a localised spatial
distribution $\Omega ({\bf x})$ can be initialised by any arbitrary initial
density field $n_e({\bf x})$ with an appropriate choice of $n_i({\bf x}) \equiv
n_e({\bf x}) - \Omega ({\bf x})$: the further evolution of $\Omega({\bf x},
t)$ in time through eq.~(\ref{e:vorticity}) will not depend on the particular
initial gyrocenter densities, but only on their difference.

In the following it is shown how finite ion temperature with $\tau_i >0$ in
magnetised plasmas leads to fundamentally different behaviour of the classical
vortex merger and co-advection problems, and thus the resulting spectral
properties of 
fully developed turbulence. In particular, the warm ion vortex merger problem
intrinsically depends on the initial gyrocenter density distribution.

For comparability with classical fluid merger problems the vorticity
is in the following initialised as a Gaussian
$\Omega({\bf x}, t_0) \sim \exp[-({\bf x}/\sigma)^2]$ of width $\sigma$.
In the gyrofluid model this may be obtained by a difference $\delta$ between
the amplitude $a$ of the electron compared to ion gyrocenter density, in the
$(x,y)$ plane perpendicular to ${\bf B}$:  
\begin{eqnarray} 
n_{e0} &=& a \exp[-((x-x_0)/\sigma)^2 - ((y-y_0)/\sigma)^2 ], \label{e:gaussne} \\
n_{i0} &=& \Gamma_1^{-1} (1-\delta \; \Gamma_0) \; n_{e0} \label{e:gaussni}
%n_{i0} &=& (1-\delta) \; n_{e0}.
\end{eqnarray}
This initialises a vorticity 
$\Omega = \Gamma_0^{-1} ( n_{e0} - \Gamma_1 n_{i0} ) = \delta \; n_{e0}$ 
for both cold and warm plasmas.
In the merger problem two vortices with the same amplitude are
placed next to each other with an initial density peak distance $\Delta_0$.

The Euler equation as a standard model for fluid flow is, like the underlying
Newtonian particle motions, invariant under parity transformation 
${\cal P}:(t,{\bf x}, {\bf v}) \rightarrow (t, -{\bf x}, -{\bf v})$.  
The evolution of flow patterns is thus symmetric with respect to simultaneous
point reflection (${\bf x} \rightarrow - {\bf x}$) and reversal
of the flow direction (${\bf v}\rightarrow - {\bf v}$). 
This implies that also for the fluid-like (cold plasma) case $\tau_i=0$ a
change in sign of the initial (pseudovector) vorticity distribution
($\Omega({\bf x}) \rightarrow -\Omega({\bf x})$) leads to a spatially
anti-symmetric evolution of the vortex merger. 

For initialisation via eqs.~(\ref{e:gaussne}, \ref{e:gaussni}), 
this can either be obtained by $n_{e0} \rightarrow - n_{e0}$ (with amplitude 
$a \rightarrow -a $), or alternatively by leaving $a$ positive but setting
$\delta \rightarrow -\delta$.  

In the cold fluid-like plasma case both choices have the same result:
in Fig.~\ref{f:merger0} such a classical merging of co-rotating vortices is 
illustrated at several times for $\tau_i=0$ with positive ($\delta=+0.01$,
top) and negative ($\delta=-0.01$, bottom) initial vorticity. 
The vorticity field $\Omega(x,y,t)$ is shown in a $(62.5 \rho_s)^2$ 
central section (corresponding to $500^2$ grid points) of the $(128 \rho_s)^2$
computational domain ($1024^2$ grid points). The initial vortex radius for
this case is $\sigma = 6 \rho_s$ with an initial peak separation $\Delta_0 = 4
\sigma$. 
The evolution is a typical example for a 2D fluid vortex merger: 
vortices orbit each other by mutual advection and after a while
(depending on initial separation) develop encircling vorticity and density
veils, and coalesce on combined advection-diffusion time scales into a
a spiraling single vortex \cite{leweke16}. The ensembles with inverted
initial vorticity evolve exactly anti-symmetrical: negative vorticity inverts
the direction of co-rotation and of the final spiral arms of the merged vortex.

\section{FLR effects on vortex merging}

It had already been noted in the seminal work of Knorr et al.~\cite{knorr88}
from 1988 on theory of FLR effects in a guiding center plasma, that in a
(reduced) exemplary simulation of vortex merging with and without FLR effects
different behaviour appeared: while for zero Larmor radius the maxima remained
separated, a coalescense had been observed for a finite Larmor radius
\cite{knorr88}.  

In the following it is shown, that such FLR effects on vortex motion and
vortex interactions in warm plasmas are generally a result of a breaking of
the axial symmetry of the vortex in presence of any spatial asymmetry in the
initial gyrocenter density distributions, which leads to FLR induced vortex
spiraling. 

However, for finite $\tau_i>0$ no ``generic'' merger scenario like in the
fluid case can be constructed: the temporal evolution of vorticity does not
only depend on the initial (for example Gaussian) distribution
$\Omega( {\bf x})$, but further on the specific gyrocenter initial
density distributions  $n_e({\bf  x})$ and $n_i ({\bf x})$ that generate this
vorticity. 

For warm plasmas the densities are not any more passively advected like in the
cold case with $d_t \Omega = 0$ from eq.~(\ref{e:vorticity}).
The gyrofluid vorticity evolution can be understood more intuitively (compare
Ref.~\cite{madsen11}) in the long wave length limit ($b^2 \ll 1$) when the
Pad\'e form of the gyrooperators is Taylor approximated: 
\begin{eqnarray}
d_t \Omega &\approx& - \frac{\tau_i}{2} d_t \nabla_{\perp}^2 (n_e + \Omega)
 + \frac{\tau_i}{2}  [\Omega, n_i]    \label{e:warmvorticity1} \\
&\approx& - \frac{\tau_i}{2}  d_t \nabla_{\perp}^2 n_e 
+ \frac{\tau_i}{2} [\Omega, n_e].   \label{e:warmvorticity2}
\end{eqnarray}
In the first line terms up to order $b^2$ are kept, and in the second line up
to $b$. For homogeneous ${\bf B}$, the identity $(\nabla_{\perp}^2 {\bf
  v}_E) \cdot \nabla_{\perp} n = [\Omega, n]$ has been used.

The first term on the right of eqs.~(\ref{e:warmvorticity1},
\ref{e:warmvorticity2}) contains the (gyroviscous cancelled) diamagnetic 
vorticity $\Omega_{d} = \nabla_{\perp}^2 p $ with $p= \tau_i n_e$, so
that the generalised vorticity $W = \Omega + \frac{1}{2} \Omega_{d}$ obeys
$d_t W \approx \frac{1}{2} [\Omega, p]$.   
This can be interpreted as an FLR induced contribution to polarisation by
advection of vorticity along isobars of pressure $p$ \cite{madsen11}. 

The Poisson bracket vanishes when the isocontours of axially symmetric
density and vorticity profiles coincide. For a deviation from exact axial
symmetry this term gives a significant contribution to the evolution
of the (generalised) vorticity.
The vortex dynamics for $\tau_i>0$ thus depends on the specific initial values
and momentary gradients of the gyrocenter densities.

%%----------------------------------------
\begin{figure} % FIG 2
%%\vspace{0.5cm}
\includegraphics[width=12.0cm]{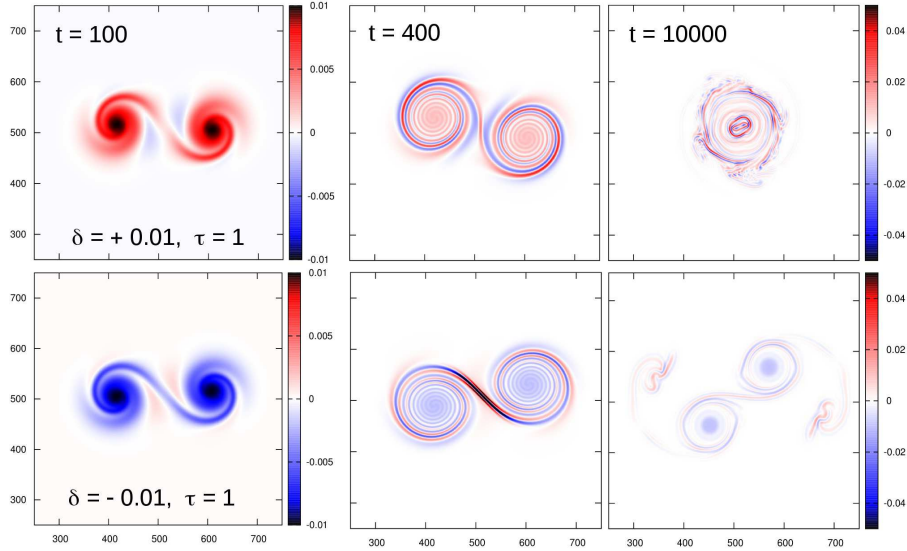}
\caption{\sl $\Omega({\bf x})$ during the co-rotating vortex pair
  interaction for $\tau_i=1$. The sign of $\delta$ as the difference between
  inital electron and ion gyrocenter density amplitudes 
  determines the fate of the pair towards merging or separation. ($t=0$ as
  in Fig.~\ref{f:merger0}.)} 
\label{f:merger1}
\end{figure}

\begin{figure} % FIG 3
%%\vspace{0.5cm}
\includegraphics[width=10.0cm]{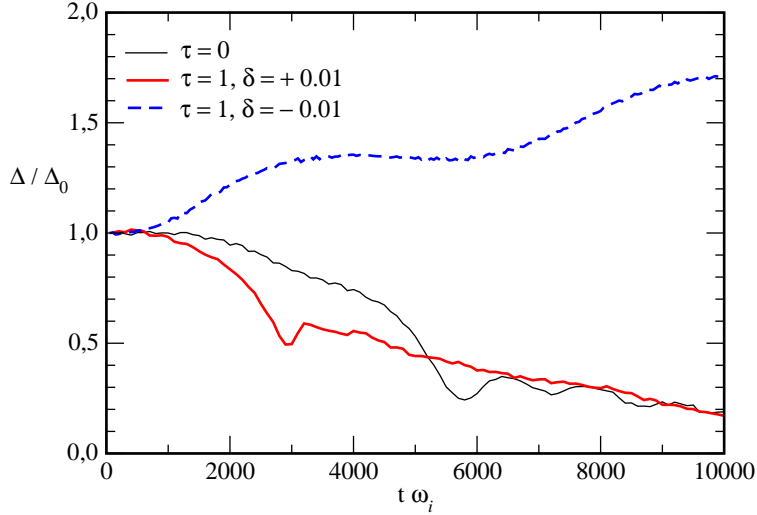}
\caption{\sl Relative vortex peak separation $\Delta/\Delta_0$ for initial
  distance $\Delta = 4 \sigma$ with $\sigma = 6 \rho_s$ for the classical fluid
case ($\tau_i = 0$, thin black line) and the FLR cases
with $\tau_i=1$, and positive (bold red line) and negative (dashed blue
line) initial vorticity relative to the magnetic field direction, respectively.} 
\label{f:delta_p4}
\end{figure}

Now the Gaussian merger is reconsidered for $\tau_i=1$. Positive
vorticity is initialised again by eqs.~(\ref{e:gaussne}, \ref{e:gaussni}) with
$a=1$ and $\delta=0.01$ and otherwise same initial conditions as above.
The evolution of $\Omega ({\bf x}, t)$ for this case is shown in
Fig.~\ref{f:merger1} (top row) for various times. 
It is observed that already in the early stages of evolution ($t=100 \;
\omega_i^{-1}$) the vortices acquire spiral arms which rapidly spin up
into a radial vorticity fine structure. 
The merging process is around twice as fast compared
to the cold plasma case. The relative separation $\Delta/\Delta_0$ between peak
densities of the vortices is plotted as a function of time in
Fig.~\ref{f:delta_p4}: 
The first minimum of distance for the warm merger is reached at around $t =
3000 \; \omega_i^{-1}$ (bold red curve) compared to the cold case at around $t =
6000 \; \omega_i^{-1}$ (thin black curve). For long times ($t > 8000 \;
\omega_i^{-1}$) in the second diffusive stage the further merging appears (at
least for the present numerical resolution and with $\nu_4 = 10^{-5}$) to
occur on similar time scales as for the cold case.

The initialisation of negative vorticity is for
$\tau_i>0$ ambigious. Negative initial density perturbation with $a=-1$ and
$\delta=0.01$ gives (similar to the cold cases of Fig.~\ref{f:merger0}) an
exacty anti-symmetric vorticity evolution: the same pattern is obtained as on
the top of Fig.~\ref{f:merger1}, with simultaneous reversal of colours (sign
of vorticity) and central point reflection. (This trivial reversed case is not
explicitly shown in Fig.~\ref{f:merger1}). Consequently the separation
$\Delta(t)$ for $a=-1$ follows the same (bold red) curve as for
$a=+1$ (with $\delta=+0.01$ for both) in Fig.~\ref{f:delta_p4}.

%%----------------------------------------

The situation completely changes when the same initial $\Omega ({\bf x}, t_0)$
is obtained by keeping $a=+1$ but setting $\delta=-0.01$ and thus changing the
relative local differences between electron and ion gyrocenter densities.
As can be seen by the second term on the right of eq.~(\ref{e:warmvorticity1}),
$n_i$ determines the FLR effect on polarisation. This means that at the
positive Gaussian (quasi-neutral) density perturbation, the ion gyrocenters
are not any more shifted outwards (as $n_i<n_e$ for $\delta >0$) but inwards
($n_i>n_e$ for $\delta<0$). This changes all gradients of $n_i$
and thus the dynamical FLR contribution to polarisation.

The effect of this relative polarisation reversal by $(\delta >0) \rightarrow
(\delta<0)$ on co-rotating vortices is significant: instead of a merger
event, a vortex separation with increasing distance is obtained. This is
illustrated as snapshots of vorticity in th ebottom row of
Fig.~\ref{f:merger1} 
and by the time evolution of the peak distances (blue dashed curve) in
Fig.~\ref{f:delta_p4}. Simulations for a range of initial separations
$\Delta_0 = 2$ to $6$ show the same general tendency, with expectedly faster
merging for shorter separations. 

\section{Dependence of vortex interactions on initial conditions}

It is not a priori clear which vorticity reversal method
is physically more relevant.
Vortices are in general not seeded, but appear dynamically mostly as a result
of the specific effects of instabilities on electron and ion densities.
An important mechanism for vorticity generation is the
drift wave instability, driven by a nonadiabatic parallel electron
response in the presence of a cross-field density gradient
\cite{horton99}. For low collisionality the relation between electron density
and potential can however often be regarded as nearly adiabatic, following
approximately a Boltzmann relation with $n_e \sim \phi$. 

We can also construct a vortex merger problem for such ``adiabatic'' vortices. 
The constraint $n_e ({\bf x}, t_0) \equiv \phi({\bf x}, t_0)$ implies that
initially $\Omega = \nabla_{\perp}^2 \phi = \nabla_{\perp}^2 n_e ~ \sim
\Omega_{d/2}$. 
In particular, setting $\phi = n_e$ in eq.~(\ref{e:polarisation}) for a
given $\phi({\bf x}, t_0)$ initialises 
$n_i \equiv \Gamma_1^{-1} [1 + \frac{1}{\tau_i}(1-\Gamma_0)] \phi$, 
which is readily evaluated in wave number space.
The choice is now whether to initialise Gaussian density/potential
``adiabatic blobs'' (which yields a shielded vorticity), or 
``adiabatic Gaussian'' vorticities.  
The latter can be achieved by inversion of a defined (Gaussian) $\Omega({\bf x},
t_0)$ to $\phi = n_e = \nabla_{\perp}^{-2} \Omega$. 

%%----------------------------------------
\begin{figure} % FIG 4
%%\vspace{0.5cm}
\includegraphics[width=10.0cm]{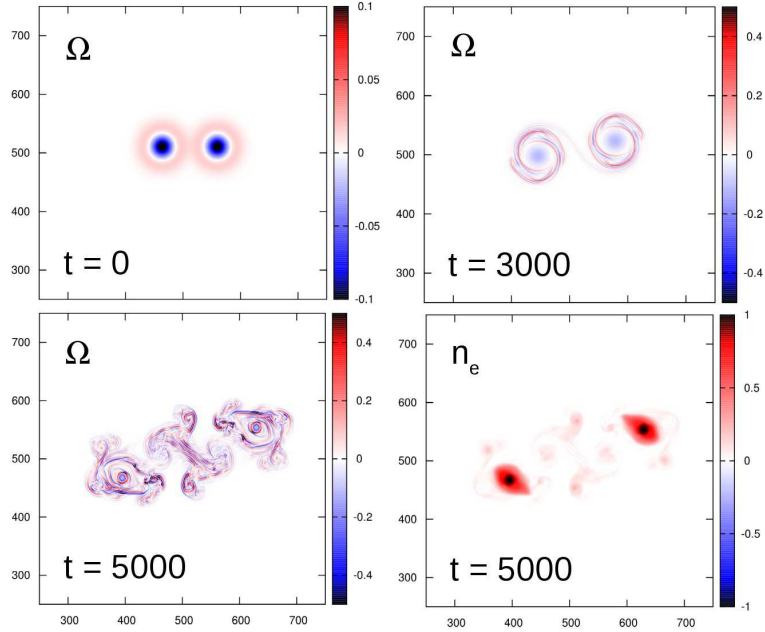}
\caption{\sl Explosive vortex repulsion out of two neighbouring ``adiabatic
  blobs'' with 
  initial Gaussian density and potential perturbations with $n_e({\bf x}, t_0) =
  \phi({\bf x}, t_0)$, which results in a shielded initial vorticity. $\Omega$
  is shown at three times, and density $n_e$ also at $t=5000 \; \omega_i^{-1}$.} 
\label{f:voradia}
\end{figure}

%%----------------------------------------
\begin{figure} % FIG 5
%%\vspace{0.5cm}
\includegraphics[width=10.0cm]{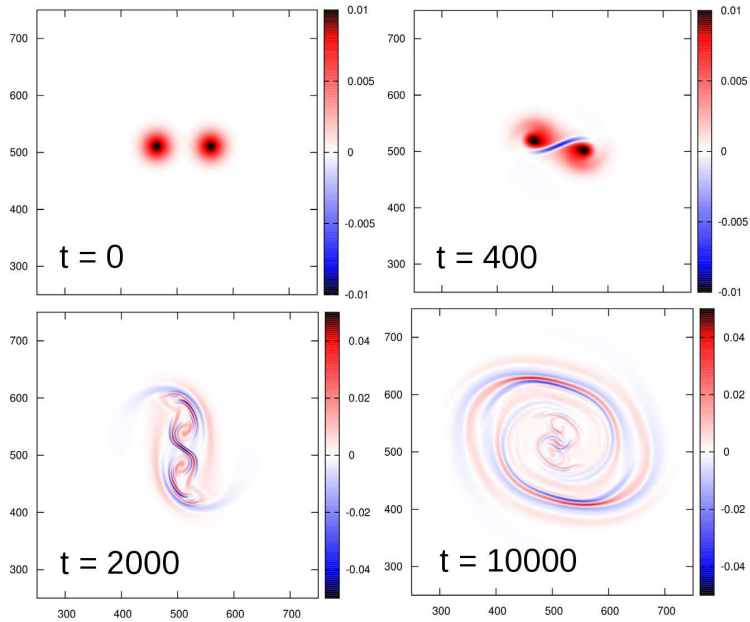}
\caption{\sl FLR accelerated merging for ``adiabatically'' ($n_e({\bf x}, t_0) =
  \phi({\bf x}, t_0)$) initialised Gaussian vorticity.} 
\label{f:adrv}
\end{figure}

%%----------------------------------------
\begin{figure} % FIG 6
%%\vspace{0.5cm}
\includegraphics[width=10.0cm]{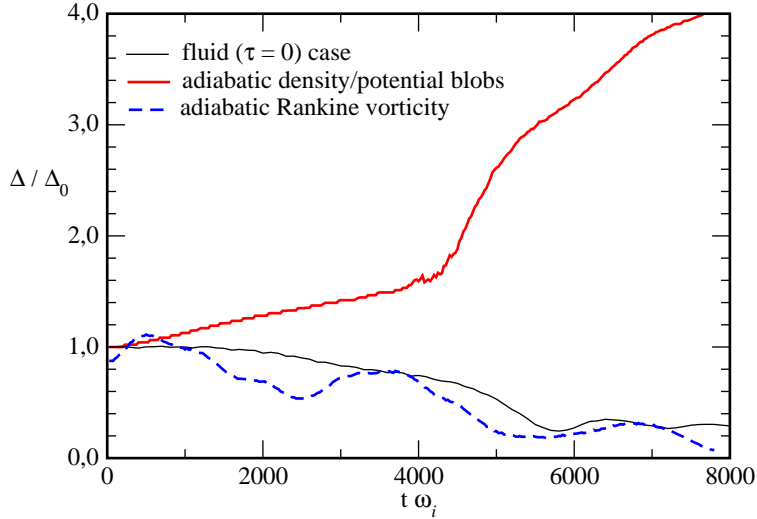}
\caption{\sl Explosive separation of ``adiabatic blobs'' (thick red curve). The
  ``adiabatic Gaussian'' case (dashed blue curve) shows again FLR induced
  acceleration of merging compared to the fluid-like (thin black curve) case.} 
\label{f:delta_ad}
\end{figure}

The ``adiabatic blobs'' case first develops small satellite vortices which are
sheared off through the shielding reversed vorticity rings around the centers. 
Subsequent collision of the satellites between the vortices rapidly
results in an explosive repulsion after formation of a vorticity tangle.
The situation is depicted in Fig.~\ref{f:voradia}.

The ``adiabatic Gaussian'' case is shown in Fig.~\ref{f:adrv}: the development
of vorticity is more similar to the fluid-like case (compare
Fig.~\ref{f:merger0}), but again with a pronounced FLR induced filamentary fine
structure and accelerated merging. 
The evolution of peak separation $\Delta$ is for both ``adiabatic'' cases
presented in Fig.~\ref{f:delta_ad}. Both cases are again anti-symmetric after
reversal of the initial amplitude.

From these examples it is obvious that the gyrofluid merger dynamics strongly
depends on the initial vorticity and density distributions, in addition to
parameters like initial relative vortex separation as in the fluid case. 

\section{Gyrospinning of asymmetric vortices}

\begin{figure} % FIG 7
%%\vspace{0.5cm}
\includegraphics[width=10.0cm]{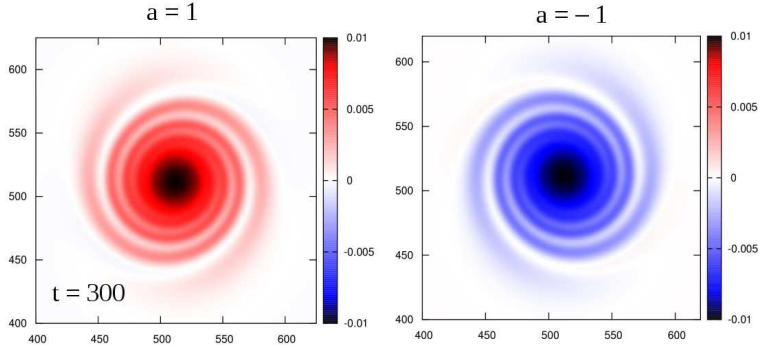}
\caption{\sl FLR spin-up of vorticity spiral arms in an asymmetric
  vortex with positive (left) and negative (right) initial 
  elongated vorticity distribution $\Omega({\bf x})$.} 
\label{f:spinner}
\end{figure}

A unique effect that is here present for all warm ion gyrofluid vortices is FLR
induced spinning. A related FLR spin-up has been observed before for the
special case of an interchange unstable (magnetic curvature driven) ``blob'',
which is characterised by the formation of a dipolar potential on top of a
monopolar (e.g. Gaussian) density or pressure perturbation 
\cite{madsen11,wiesenberger14,held16}.
While interchange ``blobs'' can acquire spinning by a range of additional 
mechanisms \cite{dippolito04,kendl15}, the FLR spin-up is in the following shown
to be a universal phenomenon and essentially a consequence of asymmetry.  
A single inviscid axially exactly symmetric vortex retains its shape.
In Fig.~\ref{f:spinner} the spin-up of FLR spiral arms is demonstrated for a
single elongated vortex ($r_y = 1.2~r_x$), initialised by
eqs.~~(\ref{e:gaussne}, \ref{e:gaussni}) with $a=+1$ (left) and $a=-1$
(right). 

%%----------------------------------------

From eq.~(\ref{e:warmvorticity2}) the FLR polarisation contribution to the
evolution of vorticity is given as $d_t \Omega \sim \frac{\tau_i}{2}
[\Omega,n]$. 
In polar $(r, \theta)$ coordinates centered on the vortex, $[\Omega,n] \sim
(\partial_r \Omega) (\partial_{\theta} n) - (\partial_{\theta} \Omega)
(\partial_r n)$ at a unit radius. For any axially symmetric $\Omega({\bf
  x}) \sim n({\bf x})$ the FLR polarisation vanishes. 

For initial distributions close to Gaussian, $\partial_r \sim (1/\sigma)$ can
be approximated.  
For symmetric  $\Omega (t_0)$ but $n=n(\theta)$, as by elongation, with
$\partial_{\theta} \sim i k_{\theta}$ a dispersion relation 
$\omega =  (\tau_i / 2 \sigma)  k_{\theta} \; \mbox{sgn}(B\cdot \Omega)$
is obtained.
%\begin{equation}
%\omega =  \frac{\tau_i}{2 \sigma}  k_{\theta} \; \mbox{sgn}(B\cdot \Omega).
%\end{equation}
A spatial density asymmetry thus spreads vorticity azimuthally with
$v_{\theta} \sim 
(\tau_i/2\sigma) \; \mbox{sgn}(B\cdot \Omega)$. Similarly, any radial density
gradient from $n(r)$ leads to radial spreading of vorticity. The combined
result is the spin-up of spiral arms as in  Fig.~\ref{f:spinner} with
orientation depending on the relative sign of magnetic field ${\bf B}$ and
vorticity ${\bf \Omega}$. 
Two neighbouring vortices mutually induce initial asymmetries similar to
elongation, resulting in rapid pre-merging spin-up.  

As a side remark, asymmetry can enter not only via non-circular vortex
initialisation, but also numerically through a coarse rectangular grid and too
close boundary proximity. Grid size and resolution have to be chosen
accordingly, that any grid spin-up artefacts evolve much slower than the
physical time scales of interest.  

%%----------------------------------------
\begin{figure} % FIG 8
%%\vspace{0.5cm}
\includegraphics[width=10.0cm]{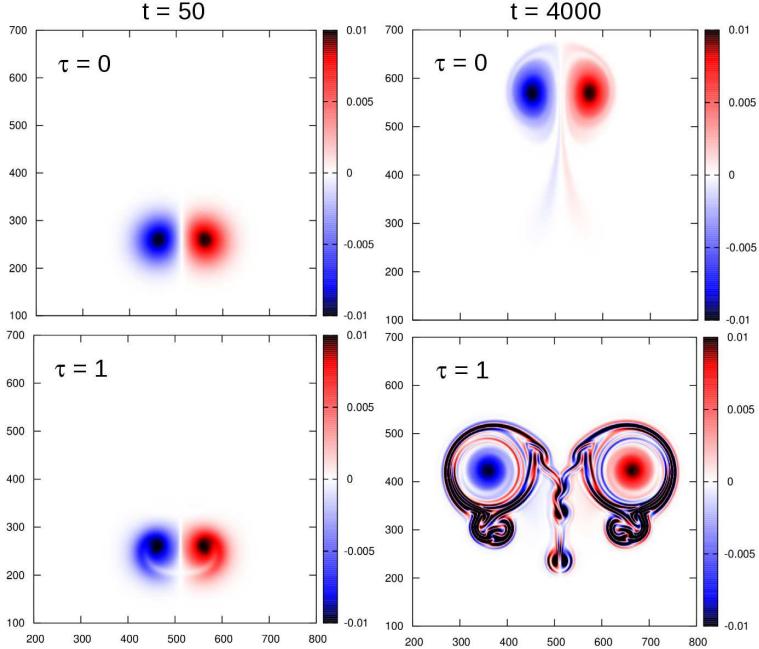}
\caption{\sl Co-advection of counter-rotating vortices at two times for the
  fluid-like ($\tau_i=0$) case (top row), and with FLR induced spin-up for
  $\tau_i=1$ (bottom row).} 
\label{f:coadvection}
\end{figure}

\section{Vorticity filamentation in vortex co-advection and turbulence}

The complementary problem to merging of co-rotating 2D vortices
is the straight co-advection of counter-rotating vortices. 
In Fig.~\ref{f:coadvection} it is shown that FLR spin-up again significantly
alters this type of vortex interaction: in a warm gyrofluid (bottom row) the
vorticity filamentation slows the joint propagation but separates the vortex
cores compared to the fluid-like cold case.

\begin{figure} % FIG 9
%%\vspace{0.5cm}
\includegraphics[width=12.0cm]{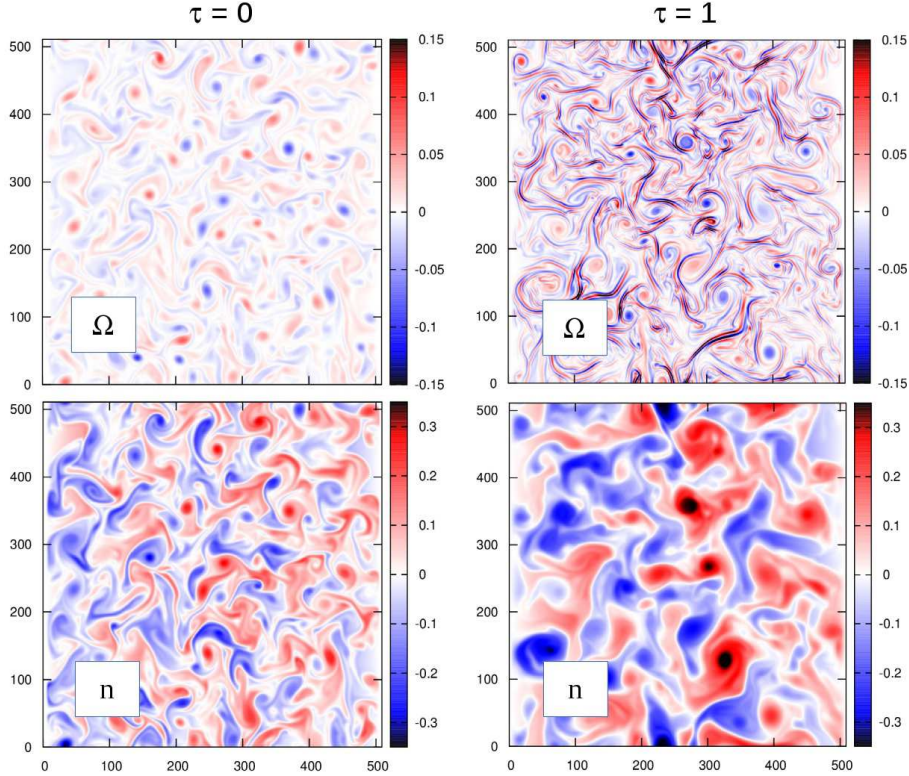}
\caption{\sl Vorticity and density fields in fully developed
  cold and warm Hasegawa-Wakatani drift wave turbulence: 
  $\Omega({\bf x})$ shows FLR-induced filamentation, and density
  fluctuations $n({\bf x})$ are enhanced on intermediate ($\rho_i k_{\perp}
  \sim 1$) scales.}  
\label{f:turbulence}
\end{figure}

In combination, merging and co-advection determine the interactions and
cascade in a turbulent sea of 2D vortices. We find that in decaying turbulence
initialised with a random distribution of density and vorticity fluctuations
analogously to eqs.~(\ref{e:warmvorticity1}, \ref{e:warmvorticity2}), the FLR
induced spinning also leads to enhanced vorticity filamentation.
Self-sustained drift-wave turbulence in inhomogeneous magnetised plasmas is in
2D effectively represented by the Hasegawa-Wakatani model \cite{hasegawa83}: 
the turbulent drive is maintained by a dissipative coupling term
$d(\phi-n_e)$ added to the right hand side of eq.~(\ref{e:continuity}) for
electrons, which emulates parallel electron dynamics for a single parallel
wave number by the parameter $d$. 
A comparison of the vorticity structure between $\tau_i=0$ and $1$ in a
saturated drift wave turbulent state for $d=0.01$ is shown in
Fig.~\ref{f:turbulence}, and demonstrates the persistence of FLR induced
vorticity filamentation in fully developed turbulence. 

Vorticity thinning has been identified as a possible explanation for the
inverse energy cascade of 2D turbulence in general fluids
\cite{chen06,bruneau07} and for drift wave turbulence \cite{manz12}, and is
here shown to be strongly enhanced by FLR spin-up in a gyrofluid. 
The amplitude of vorticity is for $\tau_i=1$ increased over the whole spectral
range, while density fluctuation amplitudes are enhanced on intermediate
($\rho_i k_{\perp} \sim 1$) scales.

%%----------------------------------------

In 3D warm gyrofluid computations of drift wave turbulence we find that the
vorticity sheets are much less pronounced but still discernible. 
The parallel connection of fluctuations along the magnetic field lines in
presence of radial zonal flow \cite{lin98} and magnetic shear \cite{kendl03}
distorts but not completely suppresses the filamentary spin-up.

\section{Conclusions}

In summary, vortex spiraling by gyroorbit effects has been shown to strongly
impact all vortex interactions in quasi-2D magnetised plasma dynamics. 
The spin-up of spiral arms was explained by an effect of density 
asymmetries on the governing drift velocities through polarisation of
gyroorbits. 
The nature and morphology of drift wave turbulence, which is of overall
importance in magnetised fusion plasmas, is essentially changed.
Gyrofluid (and gyrokinetic) simulations are able to consistently account for FLR
effects on vorticity filamentation by sufficient spatial resolution.

\bigskip

\section*{Acknowledgements}
\noindent
The author thanks M. Held (Innsbruck) for valuable discussions.

\noindent
This work was supported by the Austrian Science Fund (FWF) project Y398.


\begin{thebibliography}{00}

\bibitem{kraichnan80}
R.H. Kraichnan and D. Montgomery,
%Two-dimensional turbulence.
{\sl Rep. Prog. Phys.} {\bf 43}, 547 (1980).

\bibitem{boffetta12}
G. Boffetta and R.E. Ecke, R.E.,
%Two-dimensional turbulence.
{\sl Annu. Rev. Fluid Mech} {\bf 44}, 427 (2012).

\bibitem{horton94}
W. Horton and A. Hasegawa,
%Quasi-two-dimensional dynamics of plasmas and fluids.
{\sl Chaos} {\bf 4}, 227 (1994).

\bibitem{horton99}
W. Horton,
%Drift waves and transport
{\sl Rev. Mod. Phys.} {\bf 71}, 735 (1999).

\bibitem{scott10cpp}
B.D. Scott, A. Kendl, T. Ribeiro, 
%Nonlinear dynamics in the tokamak edge.
{\sl Contrib. Plasma Phys.} {\bf 50}, 228 (2010).

\bibitem{hahm88}
T.S. Hahm, 
%Nonlinear gyrokinetic equations for tokamak microturbulence.
{\sl Phys. Fluids} {\bf 31}, 2670 (1988).

\bibitem{brizard07}
Brizard A.J. \& Hahm, T.S.
Foundations of nonlinear gyrokinetic theory.
{\sl Rev. Mod. Phys.} {\bf 79}, 421 (2007).

\bibitem{krommes12}
Krommes, J.A.
The gyrokinetic description of microturbulence in magnetized plasmas.
{\sl Annu. Rev. Fluid Mech.} {\bf 44}, 175 (2012).

\bibitem{knorr88}
G. Knorr, et al.,
%Finite Larmor radius effects to arbitrary order.
{\sl Physica Scripta} {\bf 38}, 829 (1988)

\bibitem{dorland93}
W. Dorland and G. Hammett, 
%Gyrofluid turbulence models with kinetic effects.
{\sl Phys. Fluids B} {\bf 5}, 812 (1993).

\bibitem{beer96}
M.A. Beer and G.W. Hammett, 
%Toroidal gyrofluid equations for simulations of tokamak turbulence.
{\sl Phys. Plasmas} {\bf 3}, 4046 (1996).

\bibitem{scott05}
B. Scott, 
%Free-energy conservation in local gyrofluid models.
{\sl Phys. Plasmas} {\bf 12}, 102307 (2005).

\bibitem{scott03}
B. Scott, 
%Computation of electromagnetic turbulence and anomalous transport mechanisms
%in tokamak plasmas.
{\sl Plasma Phys. Control. Fusion} {\bf 45}, A385 (2003).

\bibitem{scott10}
B. Scott,
%Derivation via free energy conservation constraints of gyrofluid equations
%with finite-gyroradius electromagnetic nonlinearities.
{\sl Phys. Plasmas} {\bf 17}, 102306 (2010).

\bibitem{kendl15}
A. Kendl, 
%Inertial blob-hole symmetry breaking in magnetised plasma filaments.
{\sl Plasma Phys. Control. Fusion} {\bf 57}, 045012 (2015).

\bibitem{pfirsch84}
D. Pfirsch, 
%New variational formulation of Maxwell-Vlasov and guiding center theories -
%local charge and energy conservation laws.
{\sl Z. Naturforsch.} {\bf 39a}, 1 (1984).

\bibitem{brizard13}
A.J. Brizard, 
%Beyond linear gyrocenter polarization in gyrokinetic theory.
{\sl Phys. Plasmas} {\bf 20}, 092309 (2013).

\bibitem{leweke16}
T. Leweke, S. Le Dizes, and C.H.K. Williamson, 
% Dynamics and instabilities of vortex pairs.
{\sl Annu. Rev. Fluid Mech} {\bf 48}, 507 (2016).

\bibitem{madsen11}
J. Madsen, et al.,
%The influence of finite Larmor radius effects on the radial interchange
%motions of plasma filaments.
{\sl Phys. Plasmas} {\bf 18}, 112504 (2011).

\bibitem{wiesenberger14}
M. Wiesenberger, J. Madsen, A. Kendl, 
%Radial convection of finite ion temperature, high amplitude plasma blobs.
{\sl Phys. Plasmas} {\bf 21}, 092391 (2014).

\bibitem{held16}
M. Held, M. Wiesenberger, J. Madsen, A. Kendl, 
%The influence of temperature dynamics and dynamic finite ion Larmor radius
%effects on seeded high amplitude plasma blobs.
{\sl Nucl. Fusion} {\bf 56} 126005 (2016).

\bibitem{dippolito04}
D.A. D'Ippolito, J.R. Myra, D.A. Russell, G.Q. Yu, 
%Rotational stability of plasma blobs.
{\sl Phys. Plasmas} {\bf 11}, 4603 (2004).

\bibitem{hasegawa83}
A. Hasegawa, A. and M. Wakatani, 
%Plasma edge turbulence.
{\sl Phys. Plasmas} {\bf 14}, 102312 (2007).

\bibitem{chen06}
S. Chen,  et al.,
%Physical Mechanism of the Two-Dimensional Inverse Energy Cascade.
{\sl Phys. Rev. Lett.} {\bf 96} 084502 (2006).

\bibitem{bruneau07}
C.H. Bruneau, P. Fischer, H. Kellay,
%The structures responsible for the inverse energy and the forward enstrophy
%cascade in two-dimensional turbulence.
{\sl Europhysics Letters} {\bf 78}, 34002 (2007)

\bibitem{manz12}
P. Manz, G. Birkenmeier, M. Ramisch, U. Stroth, 
%A link between nonlinear self-organization and dissipation in drift-wave turbulence.
{\sl Phys. Plasmas} {\bf 19}, 082318 (2012).

\bibitem{kendl03}
A. Kendl, and B.D. Scott, 
%Magnetic Shear Damping of Dissipative Drift Wave Turbulence.
{\sl Phys. Rev. Lett.} {\bf 90}, 035006 (2003).

\bibitem{lin98}
Z. Lin, T.S. Hahm, W.W. Lee, W.M. Tang, R.B. White, 
%Turbulent Transport Reduction by Zonal Flows: Massively Parallel Simulations.
{\sl Science} {\bf 281}, 1835 (1998).



\end{thebibliography}
\end{document}